\documentclass[preprint,showpacs,preprintnumbers,amsmath,amssymb]{revtex4}

\usepackage{latexsym,amssymb,makeidx}
\usepackage{epsfig, color, ulem}
\usepackage{graphicx}
\usepackage{amsmath}

\begin{document}

\begin{center}
\rule{0cm}{2cm}
{\bf \LARGE An overview of Marchenko methods}\\
\rule{0cm}{1cm}\\
Kees Wapenaar, Myrna Staring, Joeri Brackenhoff,  Lele Zhang, Jan Thorbecke and Evert Slob 
\end{center}

\hspace{0cm}\textbf{Summary}\\
Since the introduction of the Marchenko method in geophysics, many variants have been developed. Using a compact unified notation, we review redatuming by multidimensional deconvolution and by double focusing, virtual seismology, double dereverberation and transmission-compensated Marchenko multiple elimination, and discuss the underlying assumptions, merits and limitations of these methods.

\pagebreak

\section{Introduction}

Since the introduction of the Marchenko method in geophysics \citep{Broggini2011SEG, Wapenaar2011SEG}, 
many variants have been developed, ranging from data-driven redatuming by multidimensional deconvolution
to  model-independent Marchenko multiple elimination. 
We give a brief overview of  methods developed in Delft, their underlying assumptions and their merits and limitations.

\section{The focusing function}

\begin{figure}[h]
\centerline{\epsfysize=7. cm \epsfbox{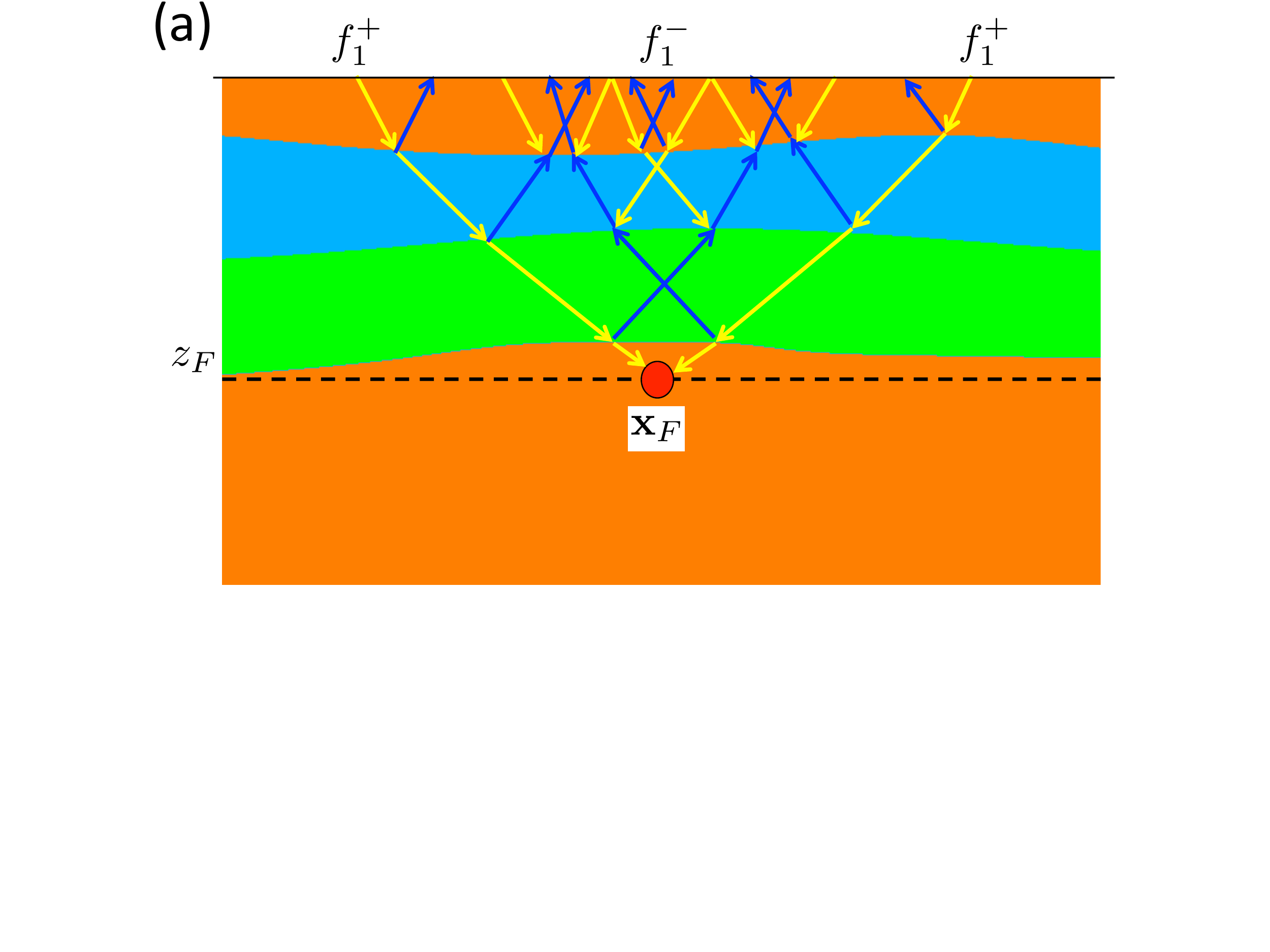}\hspace{-1cm}\epsfysize=7. cm \epsfbox{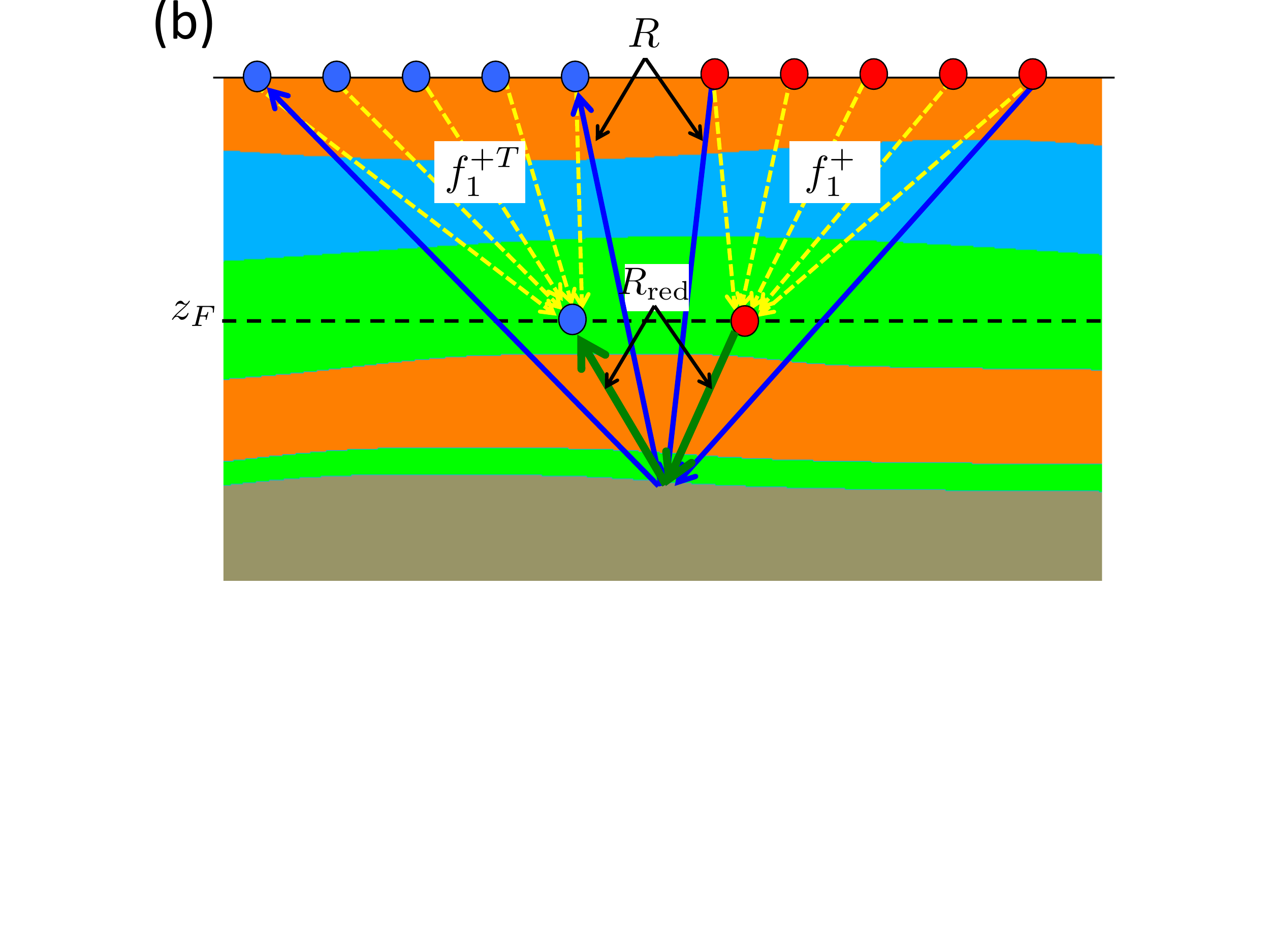}}
\vspace{-3.cm}
\caption{\small (a) Focusing function. (b) Redatuming by double focusing. The `rays' represent primaries and multiples. 
}\label{Fig1}
\end{figure}

The central concept in the Marchenko method is the focusing function \citep{Wapenaar2013PRL, Slob2014GEO}, 
which is illustrated in Figure \ref{Fig1}(a). The downgoing part of the focusing function $f_1^+$ (indicated by yellow rays), 
when emitted from the surface into a truncated version of the actual medium, focuses at a predefined location ${\bf x}_F$, without artefacts due to multiple scattering. 
The upgoing response (indicated by blue rays) is called $f_1^-$. The focusing functions can be retrieved from the reflection response $R$ at the surface and an estimate
of the direct focusing function $f_{1d}^+$ (the latter is equivalent to the standard focusing function for primaries). 
In the compact notation of \cite{Neut2015GJI}, the algorithm reads
\begin{eqnarray}\label{eq1}
f_1^+=\sum_{k=0}^\infty (\Theta R^\star\Theta R)^k f_{1d}^+,\hspace{1cm} f_1^-=\Theta R f_1^+.
\end{eqnarray}
$Rf$ stands for a multidimensional convolution of the reflection data with a function  $f$, 
the star denotes time-reversal and $\Theta$ stands for a symmetric time window $\Theta_{-t_d+\epsilon}^{t_d-\epsilon}$ that removes all events after the direct wave at $t_d$ (including the direct wave itself; $\epsilon$ is a small value to account for the finite duration of the seismic wavelet). 
The scheme requires a macro model to define the initial focusing function $f_{1d}^+$.

\section{Redatuming by multidimensional deconvolution (MDD)}

Once the focusing functions are found, the downgoing and upgoing Green's functions at the focal depth level $z_F$ follow from \citep{Wapenaar2014GEO}
\begin{eqnarray}
G^+=f_{1d}^{+\star}-\Psi R f_1^{-\star}, \hspace{1cm} G^-=\Psi R f_1^+.
\end{eqnarray}
Here $\Psi$ is the complement of time window $\Theta_{-t_d+\epsilon}^{t_d-\epsilon}$, hence, it passes the direct wave and all events after it. 
The Green's functions are mutually related via $G^-=R_{\rm red}G^+$, where $R_{\rm red}$ is the redatumed reflection response at $z_F$ of the medium 
below $z_F$. Hence, $R_{\rm red}$ follows from MDD, as follows
\begin{eqnarray}\label{eq3}
R_{\rm red}=G^-(G^+)^{-1}.
\end{eqnarray}
$R_{\rm red}$ is free of multiples related to the overburden and can be used for imaging the medium below the focal depth level. The method relies on a macro model to estimate
$f_{1d}^+$. Possible  amplitude errors in $f_{1d}^+$ are transferred to $f_1^\pm$ and $G^\pm$, but they are largely annihilated in the MDD step. A complication of the method
is that the MDD process requires a careful stabilised matrix inversion. 

\section{Redatuming by double focusing}

An alternative method to obtain $R_{\rm red}$ 
is redatuming by double focusing (Figure \ref{Fig1}(b)), formulated as 
\begin{eqnarray}\label{eq4}
R_{\rm red}=f_1^{+T}\Psi R f_1^+,
\end{eqnarray}
where superscript $T$ denotes transposition.
Equation (\ref{eq4}) is stable and can easily be applied in an adaptive way \citep{Staring2018GEO}. The retrieved response $R_{\rm red}$ contains some interactions with the overburden
and amplitude errors in $f_{1}^+$ are not annihilated.

\section{Virtual seismology}
 
The full Green's function between any two points in the subsurface can be obtained by the following variant of double focusing
\begin{eqnarray}
G=\Psi f_2^T\Psi R f_2,
\end{eqnarray}
where $f_2=f_1^+ - f_1^{-\star}$. This method can be used to forecast the response of induced earthquakes or to measure the response of 
earthquakes with virtual receivers in the subsurface \citep{Brackenhoff2019SE}.

\section{Double dereverberation} 

To reduce the sensitivity for a macro model, \cite{Neut2016GEO} proposed to project the focusing functions to the surface, according to
$v^+=f_1^+T_d$, where $v^+$ is the projected focusing function and $T_d$ is the direct arrival  of the transmission response. Since the direct focusing function is the inverse of $T_d$,
according to $f_{1d}^+T_d=\delta$ (where $\delta$ is a space-time delta function), we obtain from equation (\ref{eq1})
\begin{eqnarray}\label{eq6}
v^+=f_1^+T_d=\sum_{k=0}^\infty (\Theta R^\star\Theta R)^k \delta,\hspace{1cm} v^-=\Theta R v^+,
\end{eqnarray}
where $\Theta$ stands now for an asymmetric time window $\Theta_{\epsilon}^{t_{d2}-\epsilon}$ that removes all events at and after the two-way traveltime $t_{d2}$ of a fictitious reflector at the focal depth $z_F$. 
Since this equation does not require an estimate of $f_{1d}^+$ (unlike equation (\ref{eq1})) it is significantly less sensitive to the macro model (only $\Theta$ depends on it).
Applying $T_d^T$ and $T_d$ to the left and right of the double focusing equation (\ref{eq4}), we obtain
\begin{eqnarray}\label{eq7}
R_{\rm tar}=
T_d^TR_{\rm red}T_d=v^{+T}\Psi R v^+,
\end{eqnarray}
where $\Psi$  is now the complement of $\Theta_{\epsilon}^{t_{d2}-\epsilon}$.
The response $R_{\rm tar}$  is the redatumed response projected to the surface. It can be seen as the reflection response at the surface of the target below $z_F$ without the internal multiples related to the overburden.
Therefore the right-hand side of equation (\ref{eq7}) is a double dereverberation method \citep{Staring2020GEO}. Like double focusing, it can be applied in an adaptive way, but it is significantly less sensitive to the macro model.

\section{Transmission-compensated Marchenko multiple elimination (T-MME)} 

By replacing the asymmetric window $\Theta_{\epsilon}^{t_{d2}-\epsilon}$ in equation (\ref{eq6}) by $\Theta_{\epsilon}^{t_{d2}+\epsilon}$, 
 the event in $v^-$ at the two-way traveltime $t_{d2}$ is retained. It 
can be shown that the last event  of $v^-$ can be written as
% \citep{Zhang2019GEO}
%
\begin{eqnarray}
v_{\rm last}^-=T_d^Tr(T_d^{-1})^\star.
\end{eqnarray}
Here $r$ is the reflectivity of the deepest reflector above the focal depth $z_F$. The right-hand side can be interpreted as the primary reflection response of that reflector, 
observed at the surface and compensated for transmission losses. To obtain the complete primary reflection response, \cite{Zhang2019GEO} propose the following procedure:
apply equation (\ref{eq6}) with the modified window for all possible two-way traveltimes $t_{d2}$ (instead of focal depths $z_F$), select the sample $v^-(t=t_{d2})$ and store this to  $R_t(t_{d2})$. The resulting response $R_t(t)$ for all $t$ is the 
transmission-compensated primary reflection response at the surface. This method uses no subsurface information at all.

\section{Acknowledgements}
This research was partly funded by the European Research Council (ERC) under the European Union's Horizon 2020 research and innovation programme (grant agreement No: 742703).

\end{document}